\documentclass[aps,superscriptaddress,twocolumn,pre,longbibliography,floatfix]{revtex4-1}

\usepackage{amsmath,amsthm,amsfonts,amssymb,bm,graphicx,color,mathpazo,times}
\usepackage[colorlinks={true}, citecolor={blue}, filecolor={blue}, linkcolor={blue}, urlcolor={blue}]{hyperref}
\usepackage[caption=false]{subfig}

\usepackage{soul}

\allowdisplaybreaks

\begin{document}
	
	\title{Quantum heat engine with a quadratically coupled optomechanical system}	
	
	\author{M. Tahir Naseem}
	\affiliation{Department of Physics, Ko\c{c} University, 34450 Sariyer, Istanbul TURKEY}

	\author{\"{O}zg\"{u}r E. M\"{u}stecapl{\i}o\u{g}lu}
	\email{omustecap@ku.edu.tr}
	\affiliation{Department of Physics, Ko\c{c} University, 34450 Sariyer, Istanbul TURKEY}
	
	\date{\today}
	
\begin{abstract}
We propose a quantum heat engine based on a quadratically coupled optomechanical system. The optical component of the system is driven periodically with an incoherent thermal drive, which induces periodic oscillations in the mechanical component. 
Under the action of the quadratic optomechanical interaction, the mechanical mode evolves from an initial thermal state to a thermal-squeezed 
steady state, as verified by calculating the Wigner functions. The dynamics of the system is identified as an
effective four-stroke Otto cycle. We investigated the performance of the engine by evaluating the dissipated power, the maximum power under a load, and the maximum extractable work. It is found that the engine operating with quadratic optomechanics is more powerful than the one
operating with linear optomechanics. The effect is explained by the presence of squeezing in the quantum state of the mechanical mode.
\end{abstract}

\maketitle

\section{I.\, Introduction}\label{sec:intro}

Quantum heat engine (QHE) is a term typically used 
to describe a machine that can harness work out of thermal resources using a quantum working
substance~\cite{HE-def,Kos-HE-Rev}. QHEs has attracted much attention in the last few decades~\cite{HE1, HE2, HE3, HE4, HE5, HE6, HE7, HE8, HE9, HE10, HE11, HE12, HE13, HE14, HE15, HE16, HE17, HE18, HE19, HE20, HE21, HE22, HE23, HE24, HE25, HE26, HE27, HE28,OM-HE1, OM-HE2, OM-HE3, OM-HE4, OM-HE5} and some experimental demonstrations have been reported~\cite{Eilon2019, Ronzani2018, Maslennikov2019}. Recently, another class of quantum machines which can convert
useful energy out of non-equilibrium reservoirs, in particular squeezed thermal noise, has been
theoretically proposed~\cite{PhysRevE.93.052120} and experimentally observed~\cite{Togan2017}. While advantages of
quantum reservoirs on the engine performance are shown to be significant~\cite{Corr1, Corr2, Corr3, Corr4, Corr5, Corr6, Umit-OM}, 
it is not clear if additional 
complexity and extra energy cost of
preparing such quantum correlations would reduce their supremacy over classical resources or not. Following a more orthodox approach here,
we ask if we can see benefits of profound quantum states, in particular squeezing, of a working system to harvest work out of classical heat baths.
For that aim, we consider a system known for its capability of generating squeezing, namely quadratically coupled optomechanical
system as our working substance~\cite{gen-quad}. 

Optomechanical working substance is a natural proposal to investigate quantum thermodynamics of QHEs. It has both the
steam-like and piston-like components which are the optical and the mechanical subsystems, respectively. On the other hand, all the existing
proposals of optomechanical QHEs are limited to linear coupling~\cite{OM-HE1, OM-HE2, OM-HE3, OM-HE4, OM-HE5,Umit-OM}. Linear coupling
describes the radiation pressure induced displacement of the mechanical mode, which in terms of quantum states yield a coherent thermal
state. This state is in fact a close analog of a classical state and yields only a marginal difference if the engine harvest the classical resources
stochastically~\cite{Umit-OM}. It has been recently argued that the mechanical mode can be externally pumped with a squeezed drive. 
In Ref.~\cite{Ghosh12156}, a quantum heat engine is proposed based on a two-level system (TLS) as the working fluid that simultaneously interacts with the cold and hot baths, in addition,  it is also coupled with a cavity that plays the role of the piston. It is reported that, when the quantized piston mode is subject to a non-linear (quadratic) drive, it evolves into a thermal-squeezed state. The work capacity of the piston is considerably enhanced for the quadratic drive as compared to the case when the piston mode is subject to linear external drive.
Here, we
consider quadratic optomechanical coupling which can generate squeezing without the additional complexity or energy cost of any external squeezed drive. The key difference
between the model in Ref.~\cite{Ghosh12156} and our heat engine is that in
their model the engine takes energy from an external squeezed
drive, that generates quadratic interaction. Here, squeezing is induced by quadratic optomechanical interaction and engine takes energy from external incoherent thermal drives.

There is another scheme for a QHE based on an optomechanical system~\cite{OM-HE4}. For the work extraction from the system, the initial state of the mechanical mode needs to be in a so-called thermodynamically non-passive state~\cite{nonpass1}.  A non-passive state is a one from which work can be extracted unitarily until it becomes passive~\cite{pusz1978, Lenard1978, Allahverdyan2004, Palma2016, Brown2016, Skrzypczyk2015, Binder2015, Hovhannisyan2013, Klimovsky2013, Alicki2013, Felix2015, Niedenzu2016, Vinjanampathy2016, Goold2016, Niedenzu2018}. Non-passive states can also be regarded as quantum batteries~\cite{Alicki2013, Binder2015} or quantum flywheels~\cite{Levy2016}. The maximum amount of work that can be extracted from a non-passive state by means of a cyclic 
unitary transformation is called ergotropy~\cite{Allahverdyan2004}. On contrary, passive states have zero ergotropy. In our model, the initial state of the mechanical mode is completely passive. By computing the Wigner functions, we verify that the optical mode remains in a thermal state, and the mechanical mode evolves to a thermal-squeezed state, which has finite ergotropy. In order to compare the thermodynamic behavior of the proposed engine with the classical engine cycles, we plot the mean energy versus the  frequency of the optical mode modified by the mechanical feedback. This leads to the identification of an effective Otto engine cycle.
In addition, we compare the  linearly and quadratically coupled optomechanical QHEs~\cite{Umit-OM} using three figures of merit,
namely the maximum work capacity, the dissipated power, and the power under load. 

Rest of the paper is organized as follows. In Sec.~\ref{sec:quantumModel}, we give the quantum model based on the quadratic optomechanical coupling between the optical and the mechanical resonators. In Sec.~\ref{sec:quantumResults}, we calculate the Wigner functions for piston mode and identify the effective Otto cycle in our engine. In Sec.~\ref{sec:performance}, we calculate and compare the figure of merits of the output powers for the linear and the quadratic optomechanical coupling. Finally, we conclude our discussion in Sec.~\ref{sec:conclusions}.

\section{II.\, The Model}
\label{sec:quantumModel}

\begin{figure}[t!]
	\centering
            \includegraphics[width=7.0cm]{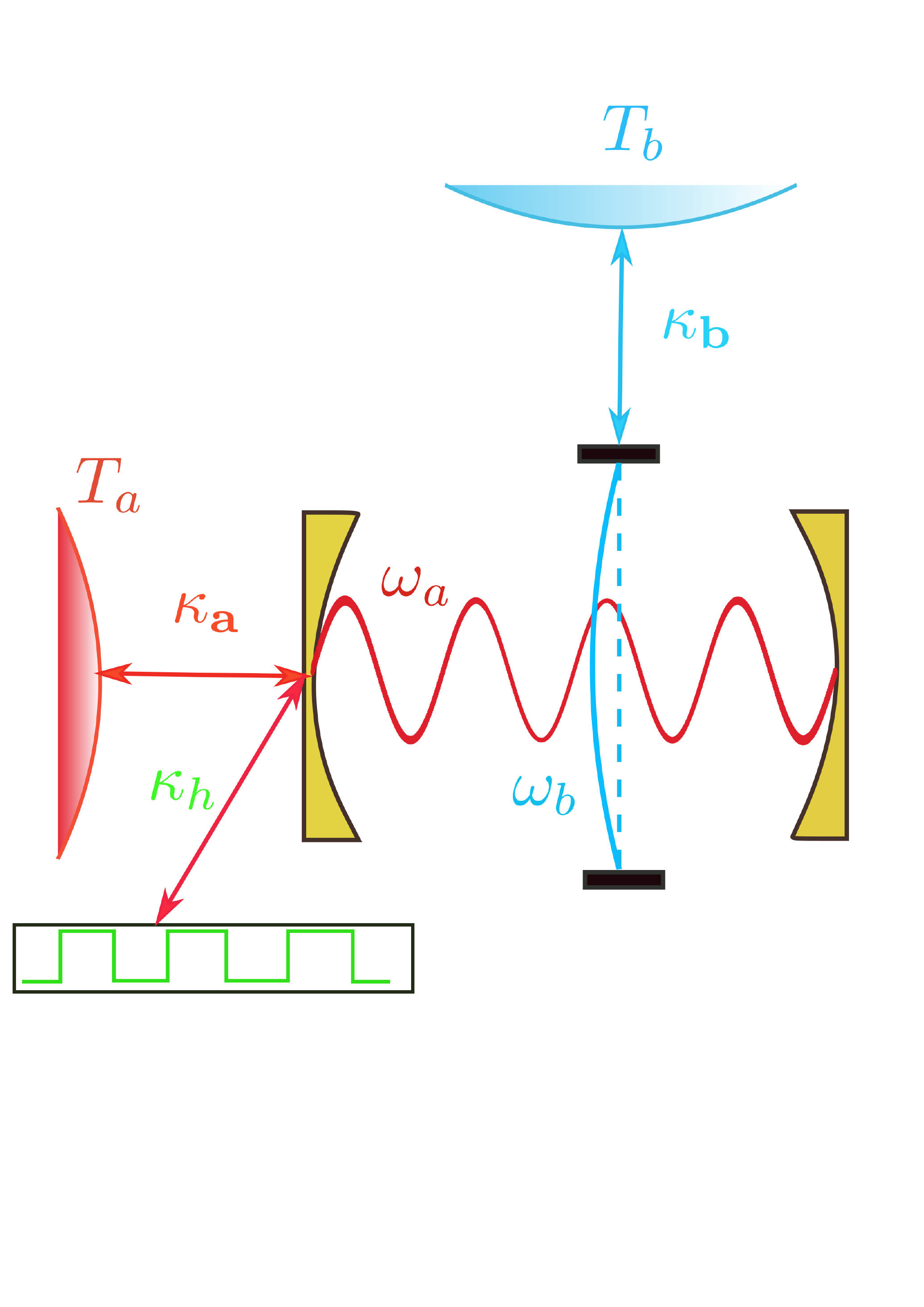}
                \caption{(Colour online)  Schematic diagram of the quantum heat engine composed of an optical cavity and a mechanical membrane that interacts via quadratic optomechanical coupling $g$. The optical resonator has the high frequency (HF)  $\omega_{\text{a}}$, and the mechanical resonator is low-frequency (LF) resonator with frequency $\omega_{\text{b}}$ . The optical (mechanical) resonator is coupled to an independent heat bath at temperature $T_\text{a}$ ($T_\text{b}$). In addition, the optical resonator is driven by a quasi-thermal periodic drive. In our heat engine cavity-mode plays the role of working fluid and mechanical resonator is our piston.}
   	\label{fig:fig1}
\end{figure}

The schematic diagram of our heat engine is shown in Fig.~\ref{fig:fig1}. It is based on an optical cavity that contains a movable membrane, and they are coupled via the quadratic optomechanical coupling. The strength of this single-photon coupling is denoted by $g$, and the optical (mechanical) resonator  has a frequency $\omega_\text{a}$ ($\omega_\text{b}$), such that $\omega_\text{a}\gg\omega_\text{b}$. We assume that the quadratic optomechanical coupling has negative values throughout this paper~\cite{Seok_2013, Seok_2014}. In our model the cavity-mode is working fluid and mechanical resonator plays the role of piston. The coupling between the optical and mechanical modes can be expressed as~\cite{NoriQuad}
\begin{equation}\label{eq:model}
\hat{H}_{\text{sys}}=\omega_\text{a}\hat{a}^{\dagger}\hat{a}+\omega_{\text b}\hat{b}^{\dagger}b+g\hat{a}^{\dagger}\hat{a}(\hat{b}+\hat{b}^{\dagger})^2,
\end{equation}  
\begin{figure*}[!t]
	\centering
	\begin{center}
		\subfloat[]{
			\includegraphics[width=4.6cm, height=4.6cm, keepaspectratio]{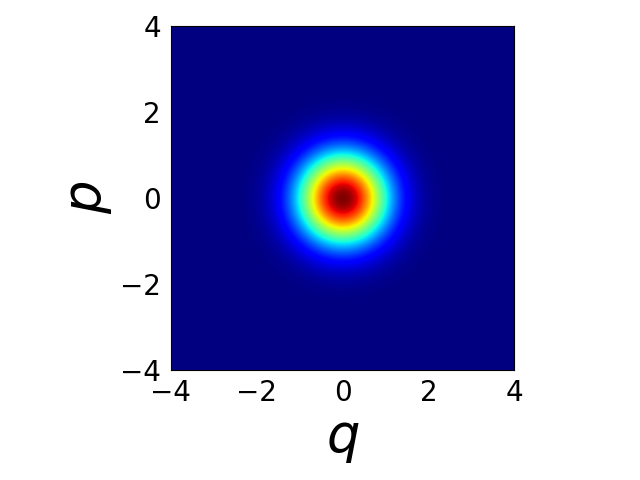}\label{fig:fig2a}
		}
		\subfloat[]{
			\includegraphics[width=4.6cm, height=4.6cm, keepaspectratio]{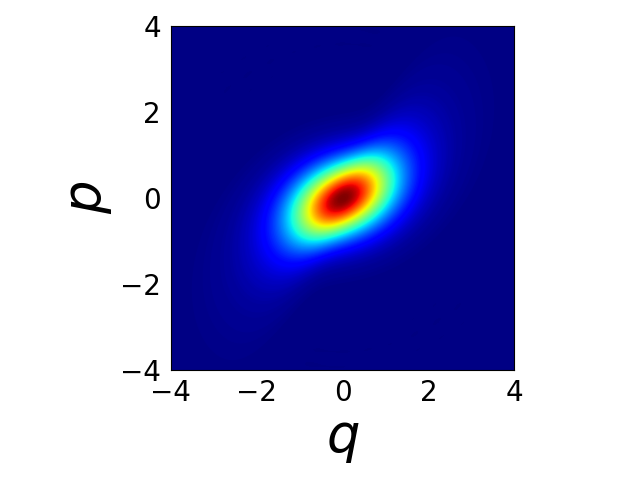}\label{fig:fig2b}
		}
		\subfloat[]{
			\includegraphics[width=4.6cm, height=4.6cm, keepaspectratio]{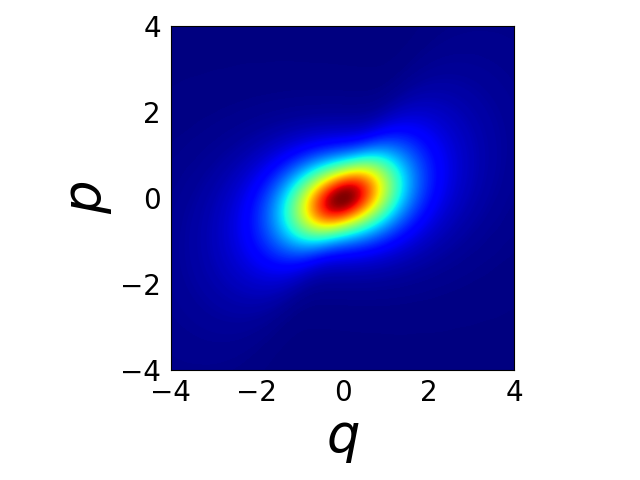}\label{fig:fig2c}
		}
		\subfloat[]{
			\includegraphics[width=4.6cm, height=4.6cm, keepaspectratio]{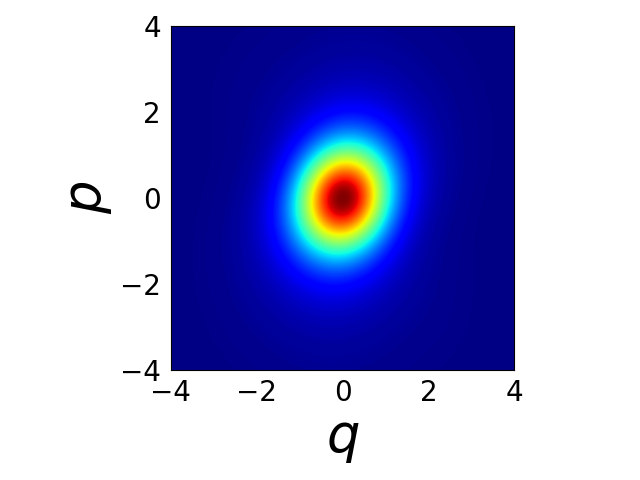}\label{fig:fig2d}
		}
	\end{center}
	\caption{(Colour online) (a)-(d) The Wigner functions in the $p,q$ field quadrature phase space of the piston mode plotted at  (a) $\omega_{\text{b}}t=0$, (b) $\omega_{\text{b}}t=30$, (c) $\omega_{\text{b}}t=300$, and (d) $\omega_{\text{b}}t=3000$ (steady-state). The other system parameters are $\kappa_{\text{a}}/\omega_{\text{b}}=4$, $\kappa_{\text{b}}/\omega_{\text{b}}=0.04$, $g/\omega_{\text{b}}=-0.6$, and $\bar{n}_{\text{h}}= 0.45$.}\label{fig:fig2}
\end{figure*}
here and in the rest of the paper we take $\hbar=1$, moreover, $\hat{a}$~($\hat{a}^{\dagger}$) and $\hat{b}$~($\hat{b}^{\dagger}$) are the annihilation (creation) operators for the optical and mechanical 
modes, respectively. We assume that the optical and mechanical resonators are coupled to two independent thermal baths at temperature $T_{\text{a}}$ and $T_{\text{b}}$, respectively. In addition, the optical resonator is driven by another a quasi-thermal periodic drive with power spectral density $S_\text{h}$. The temperatures of the thermal baths and periodic drive can be determined by,
\begin{eqnarray}
\bar{n}_\text{a}&=&\frac{1}{\exp{(\omega_\text{a}/T_\text{a})}-1},\\
\bar{n}_\text{b}&=&\frac{1}{\exp{(\omega_\text{b}/T_\text{b})}-1},\\
\bar{n}_\text{h}&=&\frac{1}{\exp{(\omega_\text{a}/T_\text{h})}-1}.
\end{eqnarray}
We take $k_\text{B}=1$. Here, $T_\text{h}$ is the temperature of periodic drive on optical resonator. We consider that $T_\text{h}>T_\text{a},T_\text{b}$.  Furthermore, $\bar{n}_\text{a}$ and $\bar{n}_\text{b}$ are the mean number of excitations in the baths for optical and mechanical resonators, respectively, while the mean number of excitations for the thermal drive at temperature $T_h$ is given by $\bar{n}_\text{h}$.
The dynamics of the system can be described by the master equation ~\cite{masterEq},
\begin{eqnarray}\label{eq:master}
\dot{\hat{\rho}}&=&-i[\hat{H}_{\text{sys}},\hat{\rho}] \\ \nonumber
&+&\kappa_\text{a}(\bar{n}_\text{a}+1)D[\hat{a}]+\kappa_\text{a}\bar{n}_\text{a}D[\hat{a}^{\dagger}]\\ \nonumber 
&+&\kappa_\text{b}(\bar{n}_\text{b}+1)D[\hat{b}]+\kappa_\text{b}\bar{n}_\text{b}D[\hat{b}^{\dagger}]\\ \nonumber 
&+&\kappa_{\text{h}}(t)(\bar{n}_{\text{h}}+1)D[\hat{a}]+\kappa_\text{h}(t)\bar{n}_{\text{h}}D[\hat{a}^{\dagger}],
\end{eqnarray}
where, $\kappa_\text{a}$ and $\kappa_\text{b}$ are coupling constants of optical and mechanical resonators with their respective thermal baths. The coupling of the optical resonator with the additional thermal drive is described by periodic time-dependent coupling coefficient $\kappa_\text{h}(t)$. $D[\hat{\alpha}]:=(1/2)(2\hat{\alpha}\hat{\rho}\hat{\alpha}^{\dagger}-\hat{\alpha}^{\dagger}\hat{\alpha}\hat{\rho}-\hat{\rho}\hat{\alpha}^{\dagger}\hat{\alpha})$ is the Lindblad dissipator superoperator with $\hat{\alpha}=\hat{a},\hat{b}$. 

\section{III.\, Nonpassivity of the Piston Mode}\label{sec:quantumResults}

The stability condition for the quadratic optomechancial system with the membrane-in-middle dictates that 	$(\omega_{\text{b}}+4\bar{n}g)>0$ ~\cite{Liao_2013}, where $\bar{n}$ is the mean number of photons inside the cavity. This requires $n_\text{h}<0.80$, in the parameters regime we consider here. In order to reduce the number of control parameters for convenience in the numerical simulations, we assume identical mean number of excitations $\bar n_\text{a} = \bar n_\text{b} := \bar n_\text{c} = 0.01$, in the thermal baths.  This is possible if $T_\text{a} / T_\text{b} = \omega_\text{a} / \omega_\text{b}$, which implies that we have $T_\text{a}\gg T_\text{b}$. We emphasize that this is not a requirement for the operation of our engine. Accordingly, the  temperatures of the baths are $T_\text{a} \sim 104$ mK and $T_\text{b}\sim 5$ mK. Thereupon,  thermal baths have the hierarchy of the temperatures $T_\text{h}>T_\text{a}>T_\text{b}$. The thermal periodic drive acting on optical resonator has a temporal profile of square wave $\kappa_\text{h}(t):=\kappa_\text{h}s(t)$.
For heating and cooling stages the square wave $s(t)=1$ and $s(t)=0$, respectively. If the external thermal pulse is on, the optical mode heats to the temperature of the thermal drive, and when the pulse is off, this mode cools to the temperature $T_{\text{a}}$. In contrast, the mechanical mode always coupled with a cold bath at temperature  $T_{\text{b}}$. Heating and cooling stage each has the same time of $\pi/\omega_\text{b}$.

Fig.~\ref{fig:fig2} shows the Wigner functions of the reduced density matrix $\rho_\text{b}=\text{Tr}_\text{a}[\rho (t)]$ of the mechanical mode at different values of scaled time $\omega_{\text{b}}t$. With $\bar n_\text{h}=0.45$, and at the start of the engine operation at $\omega_{\text{b}}t=0$, the initial state of the piston (mechanical) mode is thermal as shown in Fig.~\ref{fig:fig2a}. The state of the mechanical resonator evolves from thermal (passive) to a thermal-squeezed (non-passive) state as shown in Figs.~\ref{fig:fig2b}-\ref{fig:fig2d}. However, during the evolution, the squeezing in the piston first increases (Fig.~\ref{fig:fig2b}) and then it decreases till system reaches at steady-state (Fig.~\ref{fig:fig2d}). The decrease in the squeezing of the mechanical resonator is due to the fact that thermal noise and strong decoherence present in the system weakens the quantum correlations ~\cite{gen-quad}.

This change in the state of the mechanical resonator is due to the periodic thermal drive and quadratic optomechanical interaction. We also observed that for sufficiently small values of $\bar n_\text{h}$ the initial state of the piston mode remains thermal which has no work content. However, for higher values of $\bar n_\text{h}$ the initial state of the mechanical resonator evolves to a thermal-squeezed state. 
The Wigner functions exhibit larger widths with increasing $\bar n_\text{h}$ conforming to the increasing fluctuations. The Wigner functions are entirely positive, conforming to the fully classical dynamics associated with the mixture of thermal states character of the piston mode. On the other hand, mechanical and optical modes are still quantum correlated, which contribute to the dynamics of the $\langle \hat{n}_\text{b}\rangle$. This can be seen from the equation of motion for $\langle \hat{n}_\text{b}\rangle$ given in Appendix. We like to point out here that, if one consider linear optomechancial interaction in which the interaction term is $\hat{H}_{I}= g\hat{a}^{\dagger}\hat{a}(\hat{b}+\hat{b}^{\dagger})$, the piston (mechanical) mode evolves from thermal to a coherent-thermal (non-passive) state ~\cite{OM-HE3,Umit-OM}.
\subsection{A.\, Effective Otto engine cycle}

\begin{figure}[!t]
	\centering
	\begin{center}
		\subfloat[]{
			\label{fig:fig3a}
			\includegraphics[width=7.5cm]{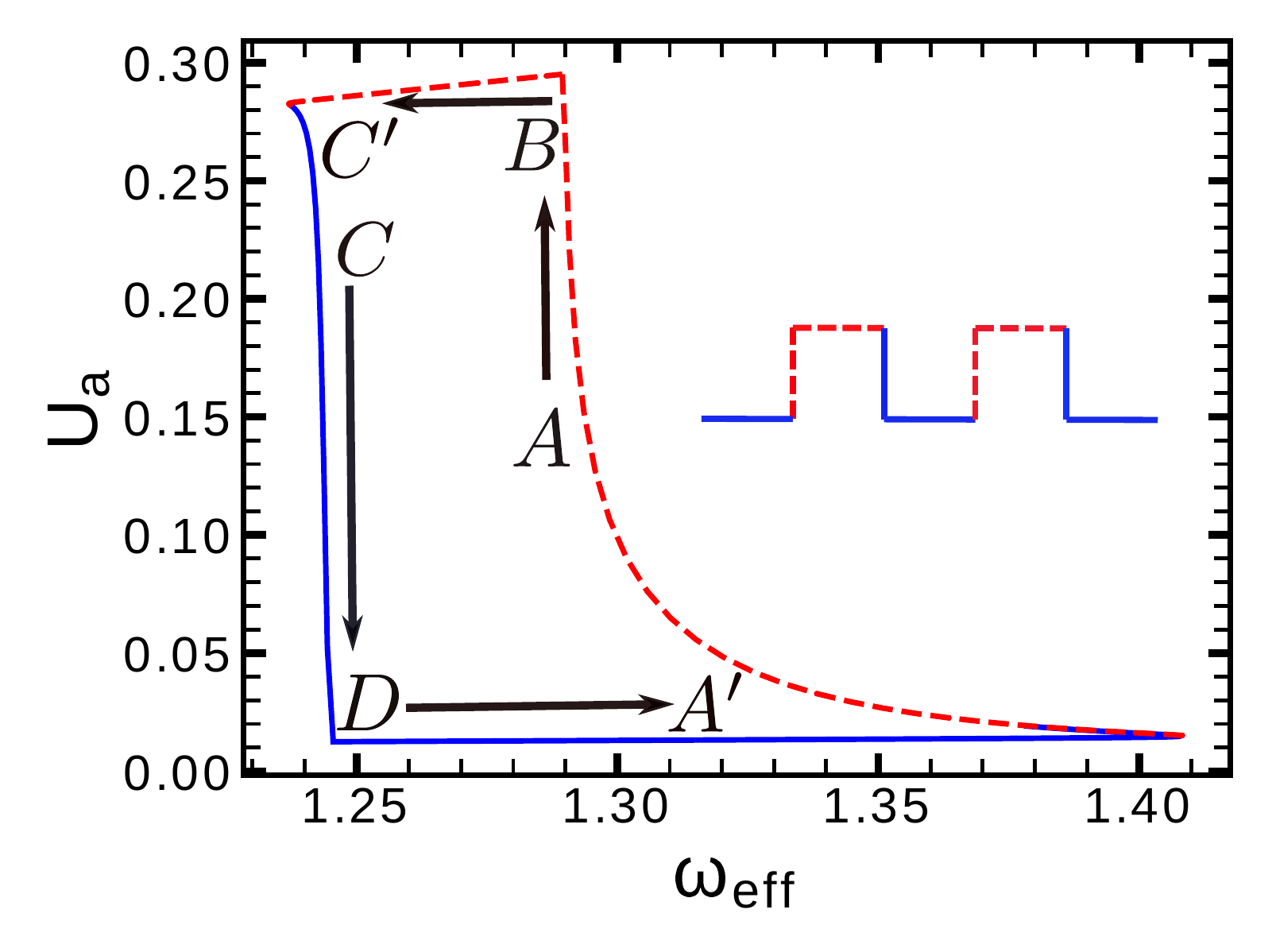}
		}\\
		\subfloat[]{
			\label{fig:fig3b}
			\includegraphics[width=7.5cm]{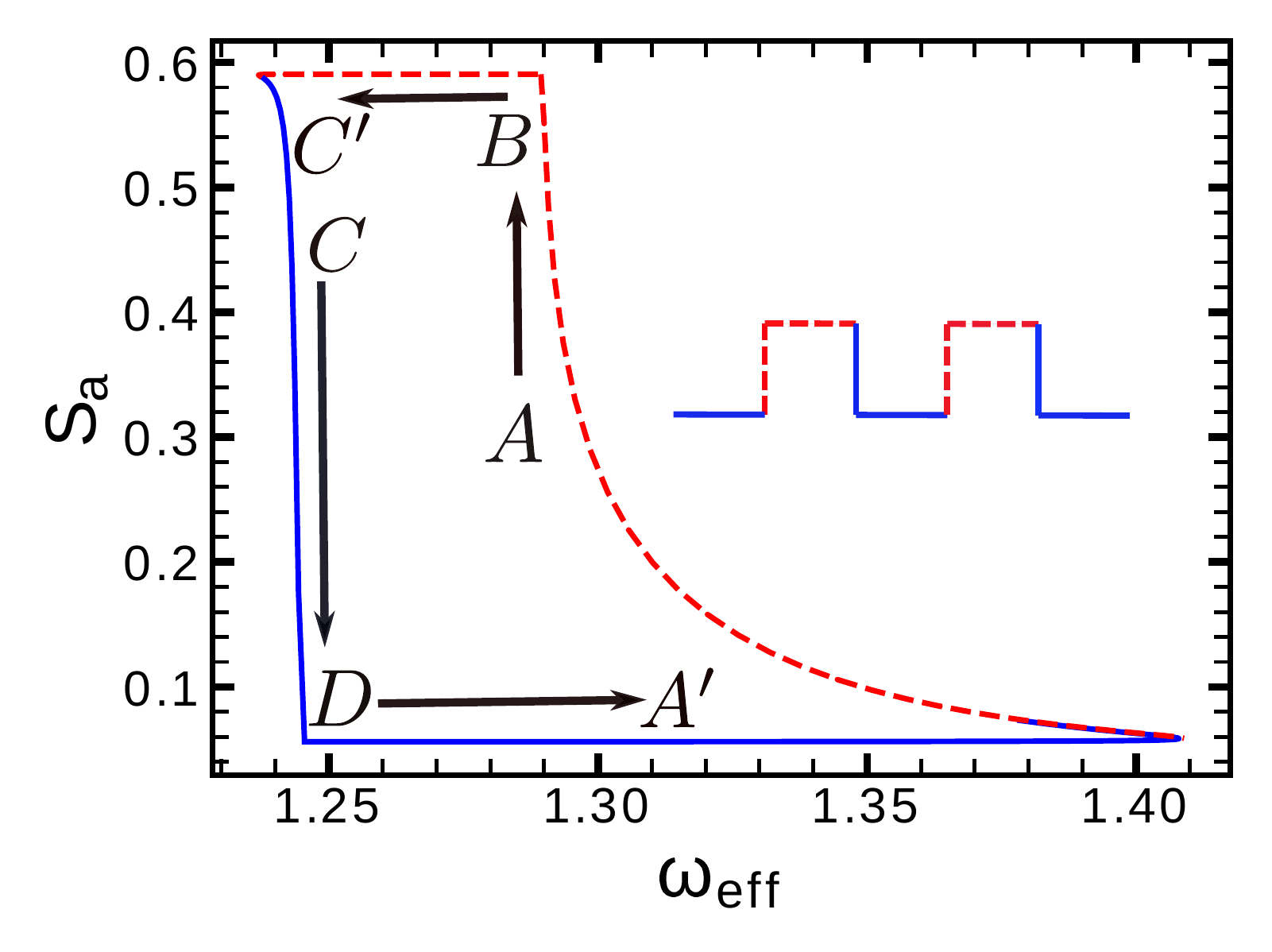}
		}
	\end{center}
	\caption{(Colour online) (a) The dependence of the mean effective energy $U_{\text{a}}=\omega_{\text{eff}}\langle \hat{n}_{\text{a}}\rangle$, and (b) the entropy $S(\rho_{\text{a}})$ versus effective frequency $\omega_\text{eff}= \omega_\text{a} + g q^2$ of the optical resonator at the steady-state. The insets shows the periodic thermal drive cycles. $\bar{n}_{\text{h}}=0.1$, and all other parameters are given in the Fig.~\ref{fig:fig2}.}
	\label{fig:fig3}
\end{figure}

To identify the effective Otto engine cycle in our model, we define the effective frequency $\omega_\text{eff}:= \omega_\text{a} + g q^2$ of the working fluid (optical resonator), where $\hat{q}=\hat{b}+\hat{b}^{\dagger}$ is the position operator of the mechanical resonator. This effective frequency can be considered as the change in the optical mode frequency associated with variations in the position of the mechanical resonator. Accordingly, we can also define the effective mean energy $U_\text{a} = \omega_\text{eff}\langle \hat{n}_\text{a}\rangle$ of the optical resonator mode, and $\hat{n}_\text{a}=\hat{a}^{\dagger}\hat{a}$ the photon number operator of the optical resonator. In the factorization of the effective energy, we have ignored the correlations between $\langle \hat{n}_\text{a}\rangle$ and $\langle\hat{q}^2\rangle$. In order to identify the engine cycle, we plot the effective mean energy $U_\text{a}$ of the working fluid against effective frequency $\omega_\text{eff}$, for $\bar n_\text{h}= 0.125$, as shown in Fig.~\ref{fig:fig3}. We like to emphasize here is that all the results presented from here onwards are at steady-state of the system unless otherwise mentioned. 

Fig.~\ref{fig:fig3a} shows a four-stage engine cycle, the first stage of our engine cycle is isochoric heating of the optical resonator under the action of the thermal pulse. Fig.~\ref{fig:fig3a} shows this stage, which is indicated by the arrow from point A to B in the figure at $\omega_\text{eff} \sim 1.292$. During this stage, the effective frequency $\omega_\text{eff}$ of mechanical resonator remains constant and the periodic thermal pulse suddenly switched on. The optical resonator receives incoherent energy and thermalizes quickly under the action of the noise pulse as compared to mechanical resonator,  which cannot follow the thermalization of the optical mode. The mean excitation number of the optical resonator reaches to steady-state $\langle \hat n_{\text{a}}\rangle^{ss}$ $\sim$ $0.0675$ for $\bar{n}_{\text{h}}=0.125$. This can be found by writing the rate equation for $\langle \hat n_{\text{a}}\rangle$ (Appendix) and finding its steady-state solution; $\langle \hat n_{\text{a}}\rangle^{ss} = ( \bar n_{\text{c}} + \bar n_{\text{h}})/2$.

The second stage in our engine cycle is the adiabatic expansion, which is indicated by the arrow from point B to C' in Fig.~\ref{fig:fig3a}. During this stage of the cycle, the thermal pulse remains active, however, due to strong dissipation in the system, there is a slight decrease in the mean effective energy $U_{a}$. The effective frequency $\omega_\text{eff}$ decreases to $\sim$ 1.245 and the displacement of the mechanical resonator from mean position increases. The entropy of the optical resonator remains constant as shown in Fig.~\ref{fig:fig3b}, the optical resonator is in thermal state and its entropy can be calculated by 
\begin{eqnarray}
S_a=(1+\langle \hat n_a\rangle)\ln{(1+\langle \hat n_a\rangle)}-\langle \hat n_a\rangle\ln{(\langle \hat n_a\rangle)}.
\end{eqnarray}
During B to C' (Fig.~\ref{fig:fig3}) of the cycle, under the action of the thermal noise pulse the state of piston starts evolving from thermal state to a thermal-squeezed state, and piston converts heat into potential energy that to be harvested. From C' to C there is a transitional stage which can not be identified with the standard thermodynamic process. The transitional stages in our effective Otto cycle are due to the presence of the inner friction in the system. The inner friction appears in quantum heat engine when the interacting part of the Hamiltonian does not commute with the non-interacting part. Due to this, the heat engine can not follow the truly adiabatic strokes in the cycle and dissipates useful energy ~\cite{Zambrini_2015}. 

The third stage of our heat engine cycle is isochoric cooling, that is indicated by the arrow from points C to D at $\omega_\text{eff}$ $\sim$ $1.245$. During this stage, the heat pulse is suddenly turned off, and the mean excitation number quickly drops to $\langle \hat n_{\text{a}}\rangle^{ss}=\bar n_{\text{c}}/2$, the optical mode still remains in a thermal state. The strong dissipation in the system also helps during the cooling stage. The last stage of the cycle is adiabatic compression of optical mode, in which the effective frequency increases to $\omega_\text{eff}$ $\sim$ $1.41$. Accordingly, the mean position of the mechanical resonator decreases, the entropy remains constant and the periodic thermal drive remains inactive during this stage.

This final stage is denoted by D to A' in Fig. ~\ref{fig:fig3b}. There is another transitional stage from A' to A that completes the engine cycle.

\section{IV.\, Performance of the engine}\label{sec:performance}
\begin{figure*}[!t]
	\centering
	\begin{center}
		\subfloat[]{
			\includegraphics[width=6.5cm, height=6.5cm, keepaspectratio]{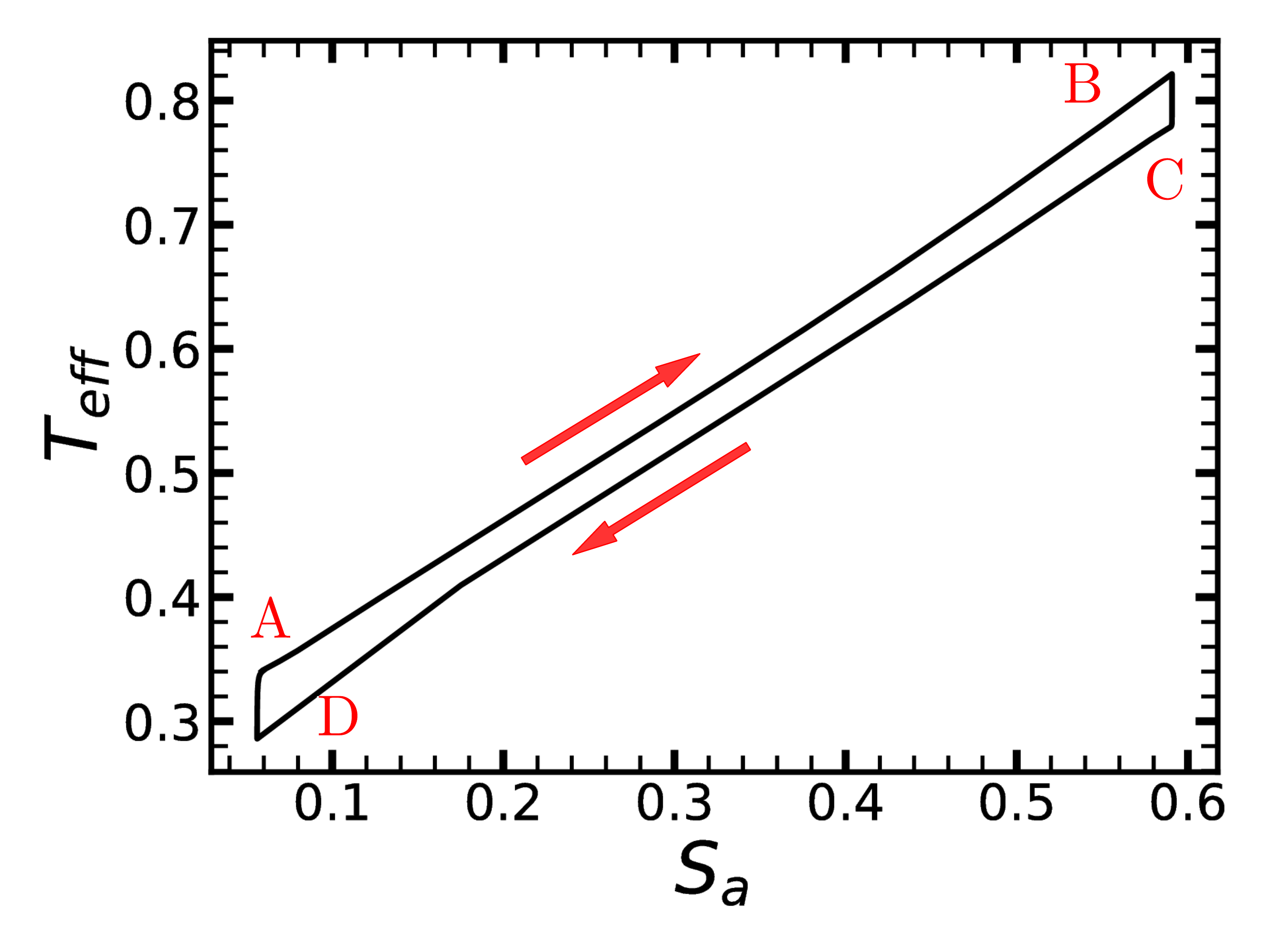}\label{fig:fig4a}
		}
		\subfloat[]{
			\includegraphics[width=6.5cm, height=6.5cm, keepaspectratio]{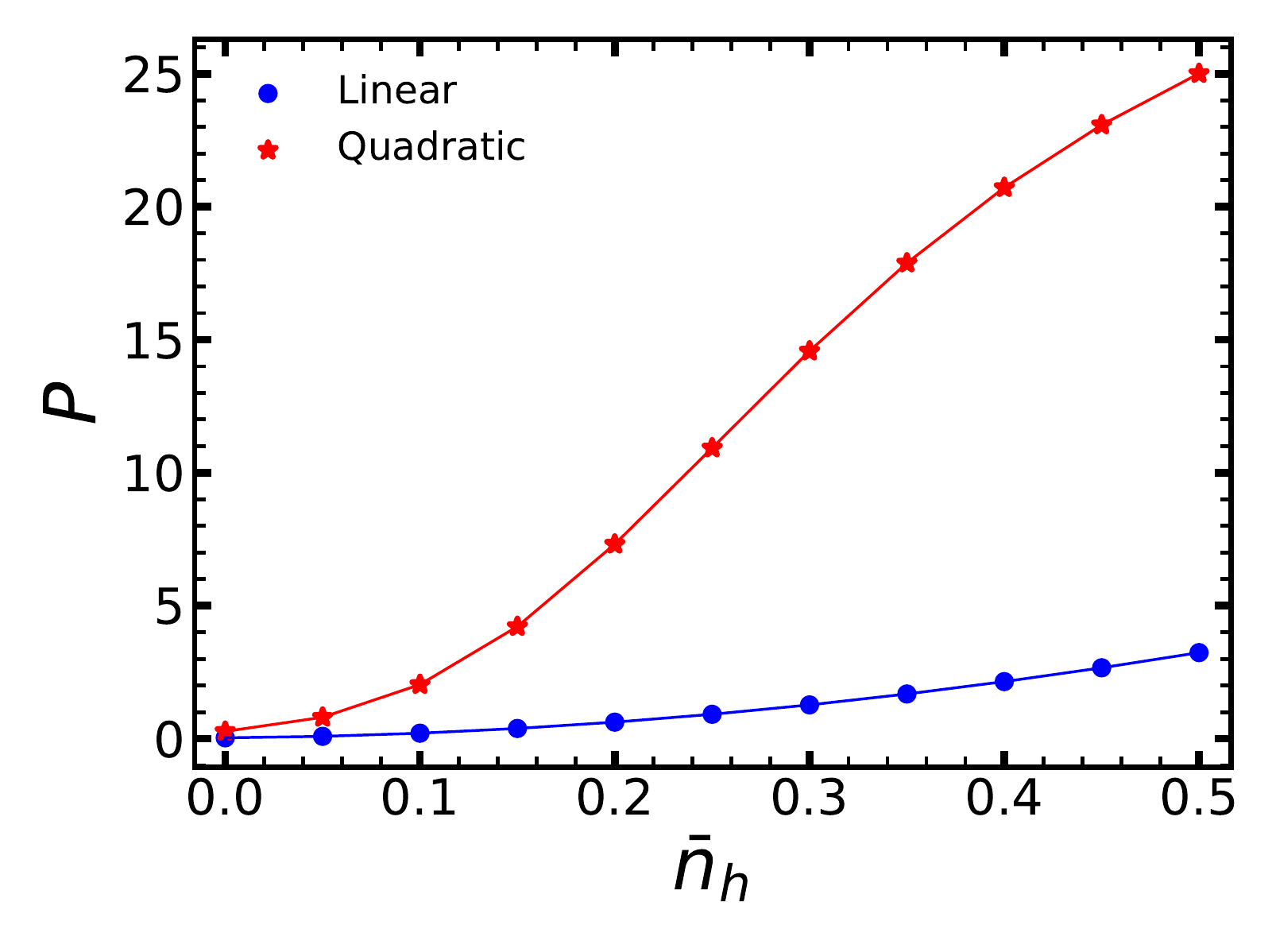}\label{fig:fig4b}
		}\\
		\subfloat[]{
			\includegraphics[width=6.5cm, height=6.5cm, keepaspectratio]{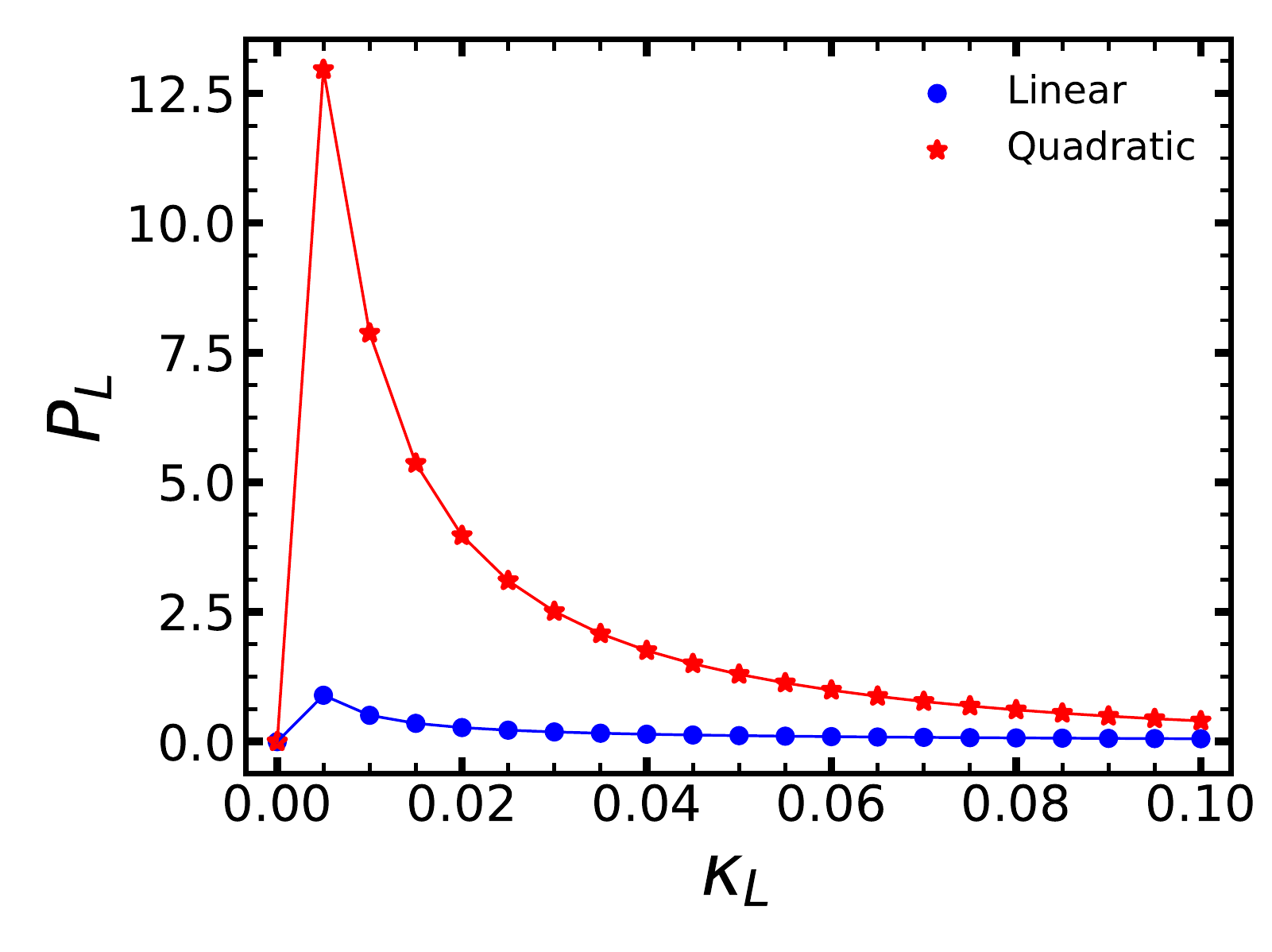}\label{fig:fig4c}
		}
		\subfloat[]{
			\includegraphics[width=6.5cm, height=6.5cm, keepaspectratio]{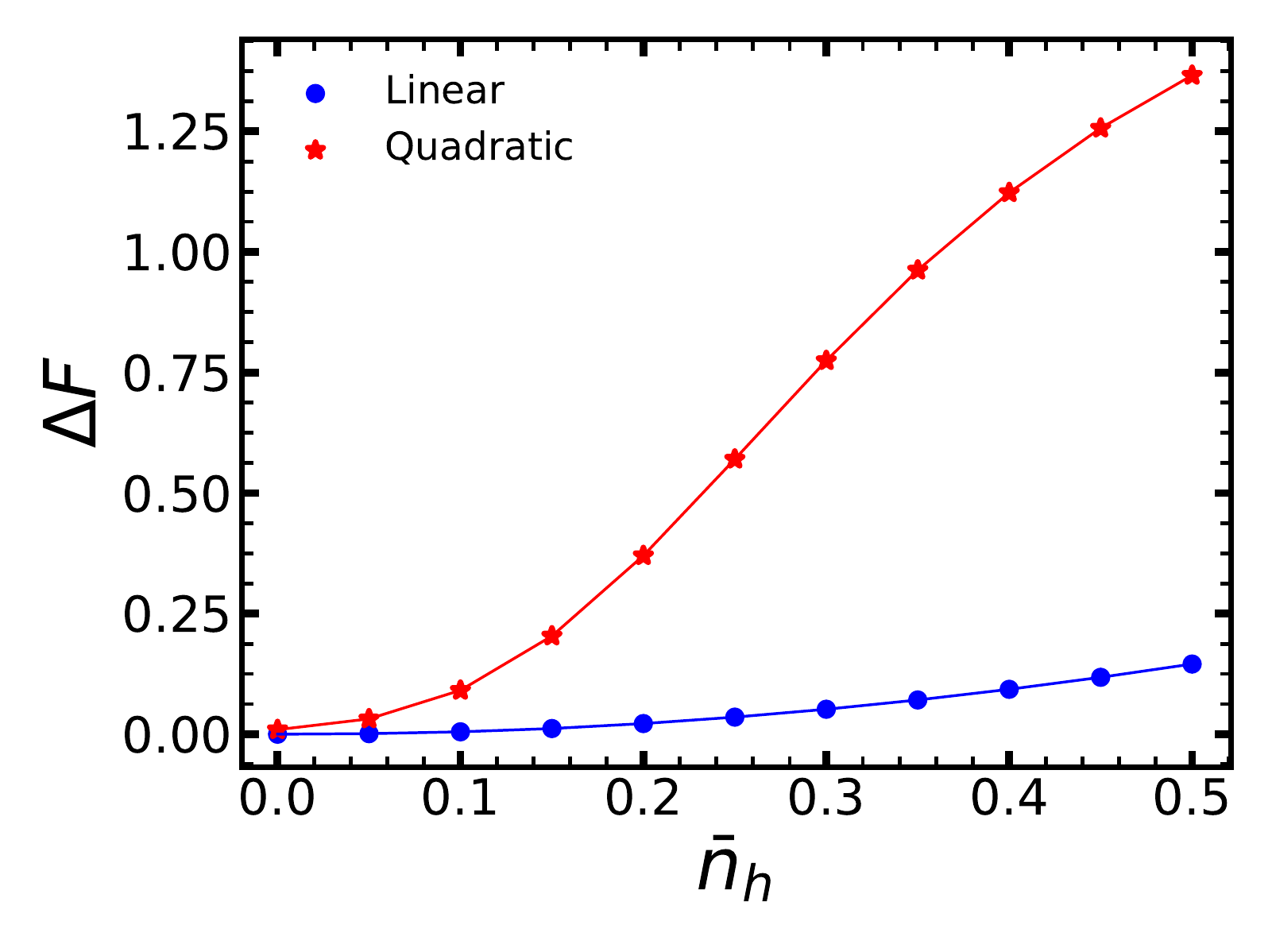}\label{fig:fig4d}
		}
	\end{center}
	\caption{(Colour online) (a) T-S diagram of the optical mode for $\bar{n}_{\text{h}} = 0.125$. 
	(b) The maximum dissipated internal power $P$ (in units of $\kappa_{b}\hbar\omega_{\text{b}}$) with respect to average number of excitations in the hot bath, (c) the maximum power under load $P_{\text{L}}$ (in units of $\kappa_{L}\hbar\omega_{\text{b}}$) as a function of external load $\kappa_\text{L}$ for $\bar{n}_{\text{h}} = 0.450$, (d) maximum extractable work $\Delta F$ (in units of $\hbar\omega_{\text{b}}$) for different values of $\bar{n}_{h}$. Blue lines are for linear and red lines are for quadratic optomechanical interaction. The values of the parameters for linear and quadratic coupling are same and given in the Fig.~\ref{fig:fig2}.}\label{fig:fig4}
\end{figure*}

Our engine can be described by the Otto-cycle excluding the transitional stages in the cycle. 
The transitional stages of the engine do not strongly affect the temperature-entropy ($T$-$S$) cycle which is shown in  Fig.~\ref{fig:fig4a}. This ($T$-$S$) cycle looks similar to standard Otto engine cycle. Here we introduce an effective temperature, as the optical resonator remains in the thermal state the effective temperature can be given by
\begin{eqnarray}
T_{\text{eff}}=\omega_{\text{eff}}/\ln{(1+1/\langle\hat n_a\rangle)}.
\end{eqnarray}
The shape of the $T$-$S$ cycle of our four-stage engine is similar to experimentally obtained cycle for single-atom heat engine proposed in ~\cite{SingleAtom}, and for nanomechanical Otto engine driven by squeezed reservoir ~\cite{OttoShape}.
 
{\it Area of $T$-$S$ cycle}: The area of the effective $T$-$S$ curve defines the useful energy content which is stored in the thermal-squeezed state of the mechanical resonator and not dissipated as heat. The piston mode undergoes to coherent oscillations due to the cyclic work output of the effective Otto engine.  This can cause ever-increasing oscillation amplitude of the piston mode, however it is balanced by the friction effect of the cold baths attached to the mechanical resonator.  The amplitude of piston mode oscillations can be determined using the methods in circuit QED~\cite{masterEq}. The area of the $T$-$S$  cycle can be considered as a figure of merit for the potential work output from the working fluid.
The shape of the cycle in Fig.~\ref{fig:fig4a} can be approximated by a trapezoid, and we can estimate the net output work by $W_a\sim2.7\times 10^{-2}\hbar\omega_a\sim 1.7\times 10^{-25}$ joules. The power can be calculated if we divide the work output with the heating pulse period $2\pi/\omega_b=2$ ns, which is the cycle time of our engine. The power from working fluid for the parameters we considered is $P_a\sim 8.9\times10^{-17}$ W.  The input heat taken by working fluid can also be calculated from the Fig.~\ref{fig:fig4a}, and we find $Q_{\text{in}}\sim 0.30\hbar\omega_a$.  Considering these values of work output and heat intake the efficiency of our engine becomes $\eta=W_a/Q_{\text{in}}\sim 0.09\%$.  We note that similar values of work output and efficiency can be found from the $\langle\hat n_a\rangle$-$\omega_{\text{eff}}$ cycle diagram, which is similar to Fig.~\ref{fig:fig3}.  Moreover, the work output increase with the increasing values of  $\bar n_h$, but we cannot increase this indefinitely. The stability condition on quadratic optomechanical model limits the indefinite increase of $\bar n_h$.

{\it Dissipated internal power (DIP)}: Alternatively, in order to estimate the power of the heat engine another figure of merit, called dissipated internal power, can be calculated, which is given by ~\cite{OM-HE3}  
\begin{eqnarray}\label{eq:intpow}
P &=& - \text{Tr}{\omega_\text{b}\hat{n}_\text{b}\kappa_\text{b}[(\bar{n}_\text{b}+1)D_{\hat{b}}[\hat{\rho}]+\bar{n}_\text{b}D_{\hat{b}^\dagger}[\hat{\rho}]]},\\ \nonumber
&=& \omega_\text{b}\kappa_\text{b}(\langle\hat{n}_\text{b}\rangle - \bar{n}_\text{c}).
\end{eqnarray}
This describes the net flux energy dissipated by the piston mode into its environment. The qualitative behavior of the DIP is the same as of the mean excitation number of the mechanical resonator which is oscillatory around some mean value. For the given system parameters, the maximum DIP has obtained at the maximum value of the mean number of excitations of the mechanical resonator;
 $P_\text{max}$ $\sim$ $\langle\hat{n}_\text{b}\rangle_\text{max}$. DIP is plotted as a function of $\bar{n}_\text{h}$ in Fig.~\ref{fig:fig4b}, for both linear and quadratic optomechanical couplings. The qualitative behavior of DIP is same for both linear and quadratic coupling, and for very small values of $\bar{n}_\text{h}$, the difference between the dissipated powers is small for the linear and quadratic models. The reason for low power at small $\bar{n}_\text{h}$ is due lack of coherence building and the small amount of squeezing of the mechanical mode in linear and quadratic coupling (cf. Fig.~\ref{fig:fig3}), respectively. As we increase $\bar{n}_\text{h}$ the squeezing in the mechanical mode increases and correspondingly dissipated power has more useful work content than the incoherent energy or heat. 
We get more dissipated power for the quadratic optomechanical coupling based heat engine, than for a linear coupling model with the same system parameters, this is shown in Fig.~\ref{fig:fig4b}. For small values of $\bar{n}_\text{h}$, the difference in the dissipated power for two models is small, but for the sufficiently high values of $\bar{n}_\text{h}$ this difference becomes pronounced.  

{\it Power under load }: Another way of estimating the power of our engine is dissipated power under load. When a load is attached to mechanical resonator, then power dissipates in its presence is termed as dissipated power under load  ($P_\text{L}$) ~\cite{OM-HE3}. The external load attached to the mechanical resonator creates additional damping along with friction introduced by the environment of the mechanical resonator. Accordingly, the dynamics of the system changes and we replace $\kappa_\text{b}\to \kappa_\text{b} + \kappa_\text{L} $. The expression for $P_\text{L}$ is similar with Eq.~(\ref{eq:intpow}) and given by
\begin{eqnarray}\label{eq:extpow}
P_{\text{L}} &=& - \text{Tr}{\omega_\text{L}\hat{n}_\text{b}\kappa_\text{L}[(\bar{n}_\text{b}+1)D_{\hat{b}}[\tilde{\rho}]+\bar{n}_\text{b}D_{\hat{b}^\dagger}[\tilde{\rho}]]},\\ \nonumber
&=& \omega_\text{b}\kappa_\text{L}(\langle\hat{n}_\text{b}\rangle - \bar{n}_\text{c}).
\end{eqnarray} 
Here $\tilde{\rho}$ is obtained by substituting $\kappa_\text{b}\to \kappa_\text{b} + \kappa_\text{L}$ in the Eq.~(\ref{eq:master}). It is clear from Eq.~(\ref{eq:extpow}) that if there is no load $\kappa_\text{L} = 0$,  or when $\kappa_\text{L}\to\infty$ (very heavy load under which thermal machine won't work), dissipated power under load becomes zero. Moreover, there is optimal value of external load $\kappa_\text{L}$ for which we get maximum power $P_\text{L}^{*} = \kappa_\text{L}^{*} P_\text{L}$. This can be used as a potential figure of merit for quantum heat engine working under an external load ~\cite{OM-HE3}.  Fig.~\ref{fig:fig4c} shows the dissipated power under load $P_\text{L}$ as a function of external load $\kappa_\text{L}$. As predicted, when there is no external load $\kappa_\text{L}$ = $0$ and for $\kappa_\text{L}$ $\to$ $\infty$, power dissipated under load is zero. There is an optimal value of load ($\kappa_{\text{L}}^{*}$) exists, that yields maximum power. Again, the power for the quadratic coupling based model is greater than linear coupling, and this difference is largest for the optimal value of load $\kappa_\text{L}^{*}$.  We also note that the squeezing in the mechanical resonator decreases as we increase the external load  $\kappa_\text{L}$, and the state of the piston mode becomes completely passive in the limit $\kappa_\text{L}$ $\to$ $\infty$.  

{\it Work capacity }: Another figure of merit for the work extraction from quantum heat engine is based on non-passivity of the piston mode. This describes the maximum extractable work from the non-equilibrium steady state of the mechanical resonator. The uper bound of the maximum extractable work for a given quantum sate $\rho$ that is subject to a given Hamiltonian $H$ can be given by ~\cite{Esposito2011, Horodecki2013},
\begin{eqnarray}\label{eq:workcap}
W^{\text{max}}\leq T S(\rho||\rho^{\text{G}}(H)),
\end{eqnarray}
where $\rho^{\text{G}}(H)$ is the Gibbs state, and $S(\rho||\rho^{\text{G}}(H))$ is the relative entropy between the quantum state $\rho$ and the Gibbs state, which can be described as: $S(\rho||\rho^{\text{G}}(H))= \text{Tr}[\rho \text{log}(\rho)-\rho \text{log}(\rho^{\text{G}})]$. We can rewrite the Eq.~(\ref{eq:workcap}) as ~\cite{OM-HE3},
\begin{eqnarray}\label{eq:FreeEng}
W^{\text{max}}\leq F(\rho)-F(\rho^{\text{G}}(H))=\Delta F.
\end{eqnarray}
Here $F(\rho)$ is the free energy of the non-equilibrium steady state. In our model, this can be determined by the difference in the mean energy of the mechanical resonator and the Von Neumann entropy of the steady state of the system; $F(\rho)= \text{Tr}[\rho\hat{H}_{\text{b}}]- K_{\text{b}}T_{\text{b}} S(\rho)$. 
In addition, $\hat{H}_{\text{b}}$ is the Hamiltonian of the mechanical resonator. 
 In our heat engine, although the initial state of the piston is thermal, however, under the action of quadratic optomechanical interaction it evolves to a thermal-squeezed state. Likewise, the optical resonator works on the piston, consequently the energy is stored as extractable work in it, and state of the piston becomes non-passive. This extractable work can be calculated by the difference in the free energies as given in Eq.~(\ref{eq:FreeEng}).  We plot this $\Delta F$ in Fig.~\ref{fig:fig4d} for both linear and quadratic optomechanical coupling based models. Again, the work capacity of the piston in case of quadratic interaction is higher than for linear coupling. The reason for this increase in the work capacity is due to the ability of the squeezed state to store work. We like to emphasize here that, in the calculation of power $P_{a}$ (Fig.~\ref{fig:fig4a}), we have ignored the correlations between the mean photon number of the optical resonator and position of the piston mode; $\langle\hat{n}_a \hat{q}^2\rangle=\langle\hat{n}_a\rangle \langle\hat{q}^2\rangle$. On contrary, this factorization is not performed while calculating the dissipated internal power, power under load and work capacity, presented in Figs.~\ref{fig:fig4b}-\ref{fig:fig4d}. In addition, we calculate the work and power of the engine at steady-state, the engine passes through a large number of transient Otto cycles to reach the steady-state. During the transient regime, the piston mode has more squeezing as shown in the Figs.~\ref{fig:fig2b} and \ref{fig:fig2c} as compared to steady-state squeezing (Fig.~\ref{fig:fig2d}). The decrease in squeezing of the piston mode is due to the presence of strong dissipation in the sideband-unresolved regime of optomechanics; $\omega_{\text{b}}\ll\kappa_{\text{a}}$. To calculate the work of a particular cycle during the transient regime, one has to perform energy measurement on the system after the completion of that cycle. This will kill the quantum correlations between the optical and mechanical components of our system~\cite{PhysRevLett.118.050601}. If we calculate the work of each cycle by using the method given in Ref.~\cite{PhysRevLett.118.050601}, the output work will be different than the calculated work at steady-state reported in Fig.~\ref{fig:fig4}. This is due to the fact that in Figs.~\ref{fig:fig4b}-\ref{fig:fig4d}, the quantum correlations are present, and these correlations will not play a role if one uses the projective measurement method presented in Ref.~\cite{PhysRevLett.118.050601}. 
 
Finally, we present some remarks about the parameters of our quadratic optomechanical heat engine for the experimental realization. The parameters regime we used $\kappa_{\text{b}}<\omega_{\text{b}}<\kappa_{\text{a}}$, can be realized in a planar silicon photonic crystal cavity~\cite{Oskar_2015}, or in circuit QED by mapping the quadratic optomechanical coupling onto the superconducting electrical circuit system~\cite{NoriQuad}. The key challenge is to realize single-photon quadratic optomechanical coupling regime in which mechanical resonator frequency becomes comparable with quadratic coupling. In our work, by taking into account the stability of the system we considered $g<\omega_{\text{b}}$ for other system parameters given in Fig.~\ref{fig:fig2}. Experimentally the single-photon strong coupling regime in the quadratic optomechanical system has not been achieved so far. However, recent experimental advances in this field may able to achieve this regime in the future.  In a planar photonic crystal cavity, the single-photon coupling strength can be enhanced from a few Hz to 1 kHz~\cite{Kalaee_2016} or several hundred kHz~\cite{Oskar_2015}. Moreover, in Ref.~\cite{PhysRevA.85.053832} the quadratic optomechanical coupling strength $g$ has been estimated in MHz regime. There are several theoretical proposals that also exploit the single-photon quadratic optomechanical coupling regime ~\cite{Liao_2013, Liao2014, PhysRevA.92.023811, PhysRevA.96.013860, PhysRevA.99.013804}.   
\section{V.\, Conclusions}\label{sec:conclusions}
In conclusion, we proposed and examined a quantum heat engine based on a general quadratic coupled optomechanical system. In our model, the working fluid mode (optical) is driven incoherently with a quasi-thermal drive, and it interacts with the quantized piston mode via quadratic optomechanical interaction. Accordingly, the piston evolves from an initial thermal state to a thermal-squeezed state. These states belong to the class of so-called thermodynamically non-passive states, as work can be extracted from such states. We verified the thermal-squeezed state of the piston mode by numerically calculating the Wigner functions.  Thermodynamical properties of the heat engine are investigated by plotting the effective mean energy for different values of effective frequency of the working fluid mode. We identified an effective Otto cycle, and estimated the extractable work by calculating the area of effective $T$-$S$ cycle diagram, which ignores the quantum correlations between the optical and mechanical components, as one figure of merit.
In addition, we calculated the internal dissipated power, dissipated power under load, and work capacity of the piston which are sensitive to quantum correlations. We reported that all these figures of merits show higher work output for the quadratic interaction relative to linear optomechanical interaction.

\section{Appendix}\label{sec:Appendix}
The equations of motions for the relevant thermodynamical observables can be determined using Eq.~(\ref{eq:master}) and given by

\begin{eqnarray}\label{Eq:rateEqs}
\frac{d}{dt}\langle\hat{n}_{a}\rangle &=& A - B\langle\hat{n}_{a}\rangle, \nonumber \\
\frac{d}{dt}\langle\hat{n}_{b}\rangle &=& \kappa_{b}(\bar{n}_{b} - \langle\hat{n}_{b}\rangle)+g\langle\hat{n}_{a}(\hat{q}\hat{p}+\hat{p}\hat{q})\rangle, \nonumber \\
\frac{d}{dt}\langle\hat{n}_{a}(\hat{q}\hat{p}+\hat{p}\hat{q})\rangle &=&  A - B\langle\hat{n}_{a}(\hat{q}\hat{p}+\hat{p}\hat{q})\rangle - (\omega_{b}- i\kappa_{b})\nonumber \\ &\times & \langle\hat{n}_{a}(\hat{q}^2-\hat{p}^2)\rangle + 8ig\langle\hat{n}_{a}^2\hat{q}^2\rangle - 4i\kappa_{b}\langle\hat{n}_{a}\hat{b}^{\dagger 2}\rangle, \nonumber \\
\frac{d}{dt}\langle\hat{n}_{a}\hat{q}^2\rangle &=& A- B\langle\hat{n}_{a}\hat{q}^2\rangle + (\omega_{b}- \frac{i}{2}\kappa_{b})\langle\hat{n}_{a}(\hat{q}\hat{p}+\hat{p}\hat{q})\rangle \nonumber \\
&+&  2\kappa_{b}(\bar{n}_{b}-\langle\hat{n}_{a}\hat{n}_{b}\rangle + \langle\hat{n}_{a}\hat{b}^{\dagger 2}\rangle), \nonumber \\
\frac{d}{dt}\langle\hat{n}_{a}\hat{p}^2\rangle &=& A- B\langle\hat{n}_{a}\hat{p}^2\rangle - (\omega_{b}+ \frac{i}{2}\kappa_{b})\langle\hat{n}_{a}(\hat{q}\hat{p}+\hat{p}\hat{q})\rangle \nonumber \\
&+& 2\kappa_{b}(\bar{n}_{b}-\langle\hat{n}_{a}\hat{n}_{b}\rangle + \langle\hat{n}_{a}\hat{b}^{\dagger 2}\rangle) \nonumber \\
&+& 4g \langle\hat{n}_{a}^2(\hat{q}\hat{p}+\hat{p}\hat{q})\rangle, \nonumber \\
\frac{d}{dt}\langle\hat{n}_{a}\hat{n}_{b}\rangle &=& A - B\langle\hat{n}_{a}\hat{n}_{b}\rangle +\kappa_{b}(\bar{n}_{b}-\langle\hat{n}_{a}\hat{n}_{b}\rangle)\nonumber \\
&+&2g\langle\hat{n}_{a}^2(\hat{q}\hat{p}+\hat{p}\hat{q})\rangle, \nonumber \\
\frac{d}{dt}\langle\hat{n}_{a}\hat{b}^{\dagger 2}\rangle &=& A + (2i(\omega_{b}+2g)-B-\kappa_{b}\bar{n}_{b})\langle\hat{n}_{a}\hat{b}^{\dagger 2}\rangle \nonumber \\
&-& 2ig(2\langle\hat{n}_{a}\hat{n}_{b}\rangle + \langle\hat{n}_{a}\rangle).
\end{eqnarray}
Where $A=\kappa_{a}\bar{n}_{a}+\kappa_{h}\bar{n}_{h}$, $B=\kappa_{a}+\kappa_{h}$, $\hat{q}=(\hat{b}+\hat{b}^{\dagger})$ and $\hat{p}=i(\hat{b}^{\dagger}-\hat{b})$. The figure of merits to evaluate the performance of the engine are presented in Figs.~\ref{fig:fig4b}-\ref{fig:fig4b}, these are numerically evaluated and depend on mean excitation $\langle\hat{n}_{b}\rangle$. During the dynamics of the heat engine, the optical and mechanical modes are quantum correlated which can be verified from the equation of motion for $\langle\hat{n}_{b}\rangle$. Although equations of motions presented in ~(\ref{Eq:rateEqs}) do not form a close set, it can still be solved by the method of weakening the correlations that allows for factorizing the optical and mechanical mode operators at the higher order of equations in the hierarchy~\cite{Bonifacio_1975, Andreev_1977}. We see that the dynamics of $\langle\hat{n}_{b}\rangle$ depends on 4-operator quantum correlations between optical and mechanical subsystems, $\langle\hat{n}_{a}\hat{q}\hat{p}\rangle$. From the hierarchy of equations of motion, it can be noted that the evolution of 4-operator quantum correlation $\langle\hat{n}_{a}\hat{q}\hat{p}\rangle$ depends on mechanical quadratic squeezing explicitly under the factorization approximation applied to 6-operator correlations such as $\langle\hat{n}_{a}^2\hat{q}^2\rangle$. We remark that similar set of equations including classical thermal white noise drives based upon Langevin dynamics can be written and the engine operation influenced with the classical correlations could be obtained. The quantum correlations in the case of the standard linear optomechanical model are stronger than classical ones and hence yielding more powerful engine when it is harvesting work in the quantum mechanical cycle instead of stochastic one ~\cite{Umit-OM}. The same conclusion, with even further enhancement by quantum squeezing, applies here, too.

\end{document}